# Magnetic Blue Phase in the Chiral Itinerant Magnet MnSi

A. Hamann[1,3], D. Lamago[1,2], Th. Wolf[1], H. v. Löhneysen[1,3], D. Reznik[1,4]

[1] *Institute of Solid State Physics, Karlsruhe Institute of Technology (KIT), D-76131 Karlsruhe, Germany*
[2] *Laboratoire Leon Brillouin, CEA-Saclay, F-91191 Gif sur Yvette Cedex, France*
[3] *Physics Institute, Karlsruhe Institute of Technology (KIT), D-76131 Karlsruhe, Germany*
[4] *Department of Physics, University of Colorado, Boulder, Colorado 80309, USA*

Chiral nematic liquid crystals sometimes form blue phases characterized by spirals twisting in different directions. By combining model calculations with neutron-scattering experiments, we show that the magnetic analogue of blue phases does form in the chiral itinerant magnet MnSi in a large part of the phase diagram. The properties of this blue phase explain a number of previously reported puzzling features of MnSi such as partial magnetic order and a two-component specific-heat and thermal-expansion anomaly at the magnetic transition.

Chiral liquid crystals are liquids with disordered molecule positions but orientations (directors) ordering into spirals. In the helical phase there is a single spiral with the directors perpendicular to the pitch axis. Blue phases, which actually have a wide range of colors due to Bragg reflections of visible light, order locally, but the pitch axis is not unique. Instead, they self-organize into double-twist cylinders with multiple pitch axes radiating away from the cylinder axis [1].

Magnetic moments in chiral magnets also tend to form spirals. In a magnetic blue phase the moment orientations, analogously to directors in liquid crystals, would also form double-twist cylinders or similar structures. Up to now their existence has not been established. Yet, a lot of evidence emerged that a chiral itinerant magnet MnSi may form a magnetic blue phase in a part of its temperature-pressure ($T$-$p$) phase diagram.

In MnSi with the cubic crystal structure B20, the lack of inversion symmetry induces a chiral Dzyaloshinsky-Moriya interaction between magnetic moments [2,3]. In the absence of an applied magnetic field $B$, long-wavelength (180 Å) helically ordered magnetic domains form at low $T$ with the pitch axes aligned along either one of the eight equivalent crystallographic <111> directions. Just above $T_c$ = 29.5 K MnSi forms unusual magnetic textures, and exhibits unconventional electrical resistivity, specific heat, and other enigmatic properties reminiscent of a blue phase [4,5,6]. These persist to the lowest investigated $T$ when $T_c$ is reduced to zero by applying hydrostatic pressure [7]. Previous attempts to explain this behavior included long-range interactions [8], soft moments [9], additional nonlinear terms [4], or higher-order terms [10,11] in the Ginzburg-Landau expansion.

By using an entirely different approach, which is similar to Monte Carlo calculations commonly performed for chiral liquid crystals [16,18,19], we show that the simplest Dzyaloshinsky-Moriya nearest-neighbor interactions are sufficient to induce a very unusual magnetic blue phase under certain conditions at finite $T$. The experimental evidence on MnSi agrees with qualitative predictions of our model.

The effective moment above $T_c$ (2.2$\mu_B$ [12]) is considerably larger than the spontaneous moment below $T_c$ (0.4$\mu_B$), indicative of a $T$-induced local moment [13]. Although MnSi is an itinerant magnet, we therefore follow the standard approach and treat the magnetic degrees of freedom as localized moments [4,8-11]. The generally accepted simplest Hamiltonian for the interaction between nearest-neighbor spins in MnSi is

$$\mathcal{H} = -\frac{1}{2N} \sum_{i=1}^{N} \left( \sum_{j(i)} (J\mathbf{s}_i \cdot \mathbf{s}_j + \mathbf{D}_{i,j} \cdot (\mathbf{s}_i \times \mathbf{s}_j)) \right) \quad (1)$$

with $|\mathbf{D}_{i,j}| = D$ and $\mathbf{D}_{i,j} = -\mathbf{D}_{j,i}$. The first and second terms in brackets denote the ferromagnetic exchange and Dzyaloshinsky-Moriya interaction, respectively. $j(i)$ indexes nearest neighbors of site $i$. $\mathbf{D}_{i,j}$ points along the vector connecting site $i$ to site $j$.

For a one-dimensional (1D) linear chain the ground state of this Hamiltonian is a helix with the propagation vector along the chain and the pitch determined by $D/J$. However, in 2D and 3D this Hamiltonian is frustrated: A helix with the propagation vector along a single direction will optimize the interactions along that direction, but not along the perpendicular directions. In MnSi a weak crystal potential locks the helix into the <111> directions.

Critical fluctuations of such a system in the continuum limit should appear uniformly distributed on the same shell in $\mathbf{Q}$ space where Bragg spots appear along <111> below $T_c$. [14] Indeed, small angle neutron scattering (SANS) experiments showed a ring of intensity, i.e., a cut through the spherical shell whose radius equals 2$\pi$ divided by the helical pitch. However, the two-component anomaly of the specific heat $C(T)$ near $T_c$ [15] consisting of a weak first-order component at the phase transition followed by a broad crossover remains unexplained. Their solution [14] and other models proposed earlier did not include topological singularities although it is well known that in chiral liquid crystals the frustration of similar interactions sometimes induces the formation of blue phases characterized by topological singularities. The existence of such singularities in the form of a skyrmion lattice has been recently demonstrated in the A-phase of MnSi, which appears close to $T_c$ in a small part of the $B$-$T$ phase diagram [4].

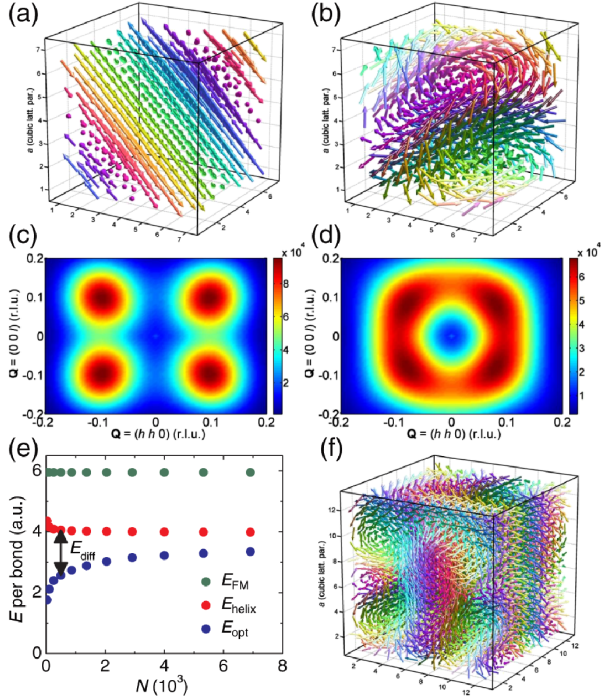

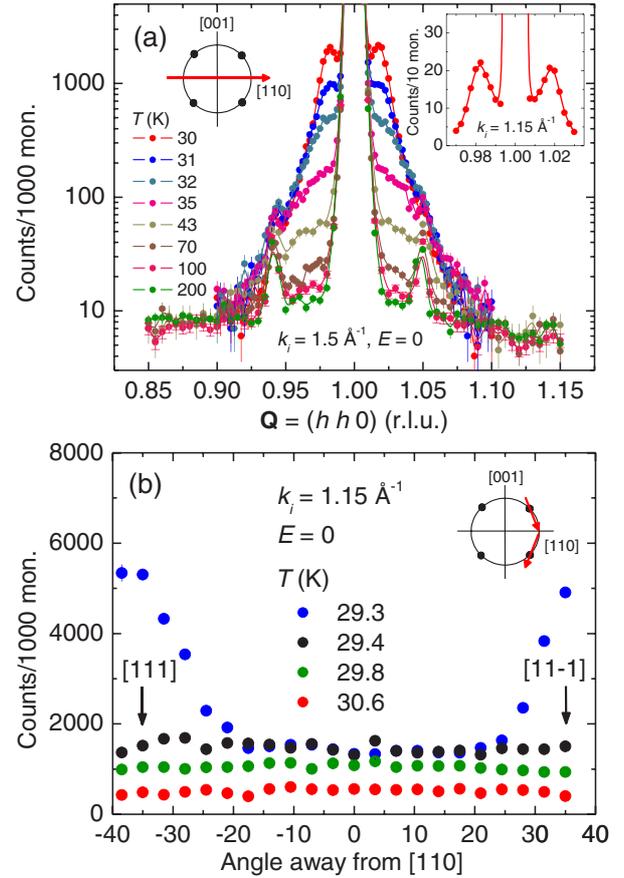

Figure 1. Calculations for clusters of spins interacting via Eq. (1). Real-space 6x6x6 unit-cell clusters: (a) helix locked in the cubic [111] direction (as in the low-$T$ phase) and (b) the optimized structure. Parallel spins have the same color. (c), (d): Calculated magnetic SANS intensity of (a) and (b) respectively. The 20% variation along the ring in (d) is most likely due to the finite size, the cubic shape, and the small number of clusters averaged over. (e) Average energies $E$ (arbitrary units) per pair of nearest-neighbor spins interacting via Eq. (1) for FM order (green), a locked helix (red), and optimized configurations (blue) vs the number of sites $N$. Zero is $E$ of a pair of spins in a fully optimized (unfrustrated) 1D chain. (f) Optimized 12x12x12 unit-cell cluster.

We performed energy-minimization calculations of magnetic moment orientations for finite-size clusters of MnSi in zero magnetic field. Such finite-size clusters capture the correct physics, because the magnetic correlation length $\xi$ in MnSi decreases on heating down to one-half of the helix period right above $T_c$ [17]. This trend continues up to at least 100 K [see Fig. 2(a)]. We model the $T$ dependence of the size of correlated domains by clusters with varying number of sites $N$ between 4 and 9000 (limited by computer power). Larger $N$ corresponds to lower $T$. Furthermore, our assumption of open boundary conditions is equivalent to magnetic domains surrounded by boundaries of randomly oriented moments. In order to ensure that the biggest clusters are larger than the helix pitch, $D/J$ was chosen deliberately to be larger than in MnSi. Different $D/J$ values gave qualitatively similar results. In real systems the crystal potential favors certain directions of the helix propagation vector, but we show below that it is weak in MnSi; thus, we first neglect it.

Figure 2. Scans across (a) and along (b) the ring of magnetic scattering at $E = 0$. Schematic insets show the scan trajectories in the [110]/[001] scattering plane. Black dots depict positions of magnetic Bragg peaks below $T_c$. Lines are guides to the eye. (a) Inset on the right: High-resolution scan across the ring revealing magnetic satellite peaks around the nuclear (110) Bragg peak. Main frame: Scans measured with relaxed resolution allowing us to trace the satellites up to 100 K (data do not change above 200 K, and thus 200 K is taken as background). Small peaks around 0.94 and 1.06 are nuclear artifacts equally present at all $T$ and hence not affecting the magnetic signal. (b) High-resolution scans along the ring from [111] to [11-1].

The calculations start with randomly oriented magnetic moments (same fixed magnitude) on a B20 lattice. Moment orientations are optimized with respect to the nearest neighbors one by one in random order until the total energy defined by Eq. (1) stabilizes. The final spin configurations obtained after each optimization routine are similar but not identical. Thus they represent not a unique ground state but local minima. Their energy $E_{opt}$ evaluated by using Eq. (1)

and normalized by the number of nearest neighbors is shown in Fig. 1(e) with energies $E_{helix}$ and $E_{FM}$ of helical and ferromagnetic (FM) order. $E_{opt}$ is nearly identical for the same cluster size and shape each time the calculation is performed and is always lower than $E_{helix}$. The energy gained from the optimization, $E_{diff} \equiv E_{helix} - E_{opt}$, is large for small $N$ (in fact, comparable to the scale of $E_{FM}$) and decreases with increasing $N$. The importance of this result will become clear below.

Figures 1(b) and (f) show resulting optimized spin arrangements. Their topology is similar to double-twist cylinders in blue phases of liquid crystals [1]. Within such a cylinder helical modulations propagate along all directions perpendicular to the cylinder axis and a line singularity of parallel moments is formed along this axis. One important difference in our structure is the presence of splay (twist away from the cylinder axis) in addition to the double-twist, leading us to the terminology of 'triple-twist'. As opposed to the blue phase of liquid crystals, where double-twist cylinders can be packed to fill 3D space, our calculations never yield a network of double-twist cylinders as a ground state. Rather, the addition of splay is essential for stacking the cylinders in a 3D network (see suppl. material Fig. S1). This structure also differs from a Skyrmion lattice, where the cylinder axes align parallel to an applied magnetic field [4,20]. However, our the triple-twist structure is similar to an amorphous network of Skyrmions proposed previously [9] but whose stability has never been proven theoretically. In order to connect with experiments, we calculated the magnetic neutron intensity following Eq. (7.61) in Ref. [21]. To mimic scattering from an infinite number of clusters, we summed the signal of several clusters of slightly different sizes in the plane spanned by reciprocal [110]/[001] axes and averaged crystallographically equivalent **Q** points. Helical order along <111> (occurring below $T_c$) (Fig. 1a) results in four diffraction spots [Fig. 1(c)], whereas the blue phase (Fig.1b) yields the ring [Fig. 1(d)] observed by SANS [14].

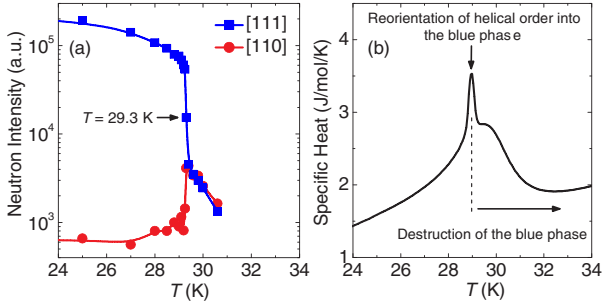

Figure 3. $T$ dependence of neutron intensity at [111] and [110] satellite positions (a) compared to the specific-heat data (b) from Ref. [15].

A moderately large crystal anisotropy stabilizes helical order along the favored direction [Fig. 1(a)] over the blue phase lacking a pitch axis with a well-defined direction [Fig. 1(b)]. Small anisotropy energy will lead to phase competition driven by the cluster size $N$. Below $T_c$ and at zero magnetic field the pitch axes of helical order are pinned along <111> [22]. However, a field as low as 0.1 T applied along a random direction overcomes the pinning potential (represented by $E_{111}$) aligning the pitch axes along that direction and it inducing a small net ferromagnetic moment [23,24]. A much larger field of 0.55 T is necessary to completely (ferromagnetically) align the moments parallel to it [23]. Comparing the two field values yields $E_{111} \approx 0.2 \, E_{FM}$. For a ratio $D/J$ that produces a pitch of about 52 Å, we calculate $E_{diff}(N) \approx 0.1 \, E_{FM}$ and 0.3 $E_{FM}$ for the largest and small clusters, respectively [Fig. 1(e)]. Thus $E_{diff}$ is of the same magnitude as $E_{111}$. At low $T$ where the correlated domains are large, $E_{111} > E_{diff}(N)$. Increasing $T$ decreases the domain size and $E_{diff}(N)$ rises because $N$ decreases. The phase transition to the blue phase occurs when $E_{diff}(N) = E_{111}$. It should be of first order because the topology of the triple twist differs from that of the simple helix. Phase transitions of blue phases in liquid crystals are also of first order. On further heating the transition to the high-$T$ paramagnetic phase should be gradual, because the stability of the blue phase given by $E_{diff}(N)$ increases with decreasing $\xi$ modeled by $N$. At low $T$ the competition with the blue phase is probably responsible for the large mosaic spread of the helical propagation vector and the strong hysteresis under applied magnetic field [25].

We tested our model with neutron-scattering experiments. Figure 2(a) shows elastic **Q** scans above $T_c$ *across* the ring observed in SANS along [110] (inset on the left). Intense magnetic satellites due to this ring are clearly visible up to 100 K on both sides of the central nuclear (110) Bragg peak. They are much broader than the **Q** resolution (0.006 r.l.u.). Figure 2(b) shows scans *along* the ring in the vicinity of $T_c$. Figure 3(a) shows the $T$ dependence of neutron intensity on the ring along [111] and [110]. Together with Fig. 2(b), it illustrates the first-order fashion of the phase transition at $T_c$ = 29.3 K (also reported in [17]): Peaks at <111> disappear on heating from 29.3 K (8% of their low-$T$ intensity) to 29.4 K, which indicates a sudden reorientation from low-$T$ helical order (intensity concentrated in <111> spots) to the blue phase above $T_c$ (intensity uniform on the ring, and hence equal at <111> and <110>). Further heating gradually reduces the intensity suggesting a crossover to the high-$T$ paramagnetic phase which naturally comes out of our calculations: The stability of the triple-twist clusters increases as $\xi$ decreases with increasing $T$ [Fig. 1(e)]. We can also understand the two-component character of $C(T)$ [Fig. 3(b)] and ultrasound attenuation up to now considered to be enigmatic [15,26]: The first-order transition has small thermodynamic weight because of the small difference in free energy of the helical phase and the blue phase. Most of the weight appears in the broad shoulder at higher $T$ due to the gradual melting of the triple-twist clusters.

Magnetic fluctuations above $T_c$ are dynamic [17,27,28] and chiral [17,29]. These, however, were measured either away from the region in (**Q**-$E$) space where we find the signature of the blue phase, or only close to $T_c$. Extending this work, we traced this signature in **Q** [Fig. 2(a)] and $E$ (Fig. 4) *on* the ring *and* up to 100 K. The result is that even well above $T_c$ the $E$ width is more than 5 times smaller than *off* the ring [28], and the **Q** width is narrower at $E = 0$ than at 0.5 meV [29] by a similar amount. The linear $E$ width in $T$-$T_c$ (Fig. 4) is qualitatively consistent with our model: increasing $T$ gradually breaks up triple-twist clusters allowing their diffusion to speed up. Remarkably, assuming simple Moriya-Kawabata fluctuations for weak itinerant FM [30] (as opposed to *chiral* fluctuations in MnSi) yields a similar $T$ dependence.

Hydrostatic $p$ reduces $T_c$, driving it to zero at 14.6 kbar where the $T$ dependence of the electrical resistivity abruptly changes from a $T^2$ (Fermi-liquid) to a $T^{1.5}$ non-Fermi-liquid (NFL) power law persisting over three decades in $T$ [31,32].

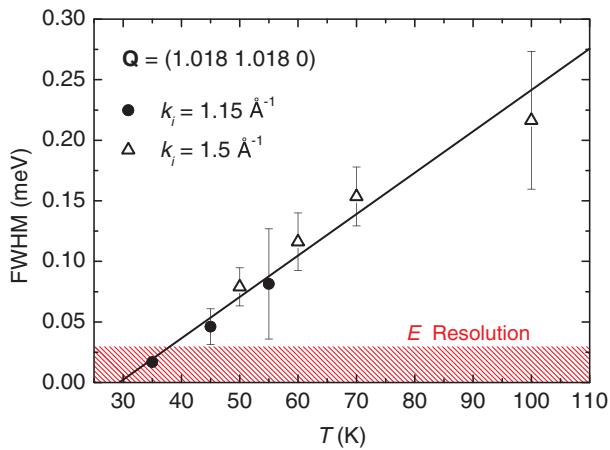

Figure 4. Intrinsic $E$ width of the magnetic signal at **Q** = (1.018, 1.018, 0) at different $T$ with a linear fit. It scales inversely with the lifetime of the correlations. Values were extracted from fitting the convolution of a Gaussian (measured experimental resolution) and a Lorentzian (intrinsic signal) to $E$ scans.

In the same $p$ range, so-called partial order (PO) appears with a diffraction signature similar to that of the blue phase [7]. The caveat is that the angular distribution is maximal along <110> while it is isotropic in the blue phase [Figs. 2(b), 3(a)]. Magnetic order affects the electronic scattering rate [31], suggesting that PO and the NFL resistivity might be related. [33] Yet, the $T^{1.5}$ resistivity was reported in a much larger part of the $T$-$p$ phase diagram than PO [32]. Our analysis points a way to resolving this paradox: The gradual crossover from the blue phase to the paramagnetic phase makes PO undetectable under $p$ once $\xi$ becomes short, since the pressure cell reduces the sensitivity compared to our experiment by two orders of magnitude due to additional background and reduced sample volume. Hence PO may appear in the same part of the phase diagram as the NFL resistivity. Finally, a skyrmion lattice in the A-phase of MnSi [4] qualitatively agrees with our model: The A-phase exists in a small $B$-$T$ window just below $T_c$ where field alignment of triple-twist cylinders is easily possible.

In conclusion, we have provided compelling evidence that the magnetic analogue of a blue phase does form in MnSi and exists at least up to 3 $T_c$. An interesting remaining issue is that of what drives the fully "isotropic" blue phase to the PO phase with increasing $p$. More detailed calculations of larger clusters including entropy to obtain thermodynamic properties are highly desirable. The elucidate the understanding of the relationship between the blue phase and NFL behavior.

The authors would like to thank L. Pintschovius, Y. Uemura, R. Hott, P. Wölfle, A. Rosch, M. Vojta, C. Meingast, N. Clark, M. Hemerle, L. Radzihovsky, and M. Lee for helpful discussions. We thank D. Petitgrand, Y. Sidis and G. Chaboussant for technical help.

**Supplementary Material**

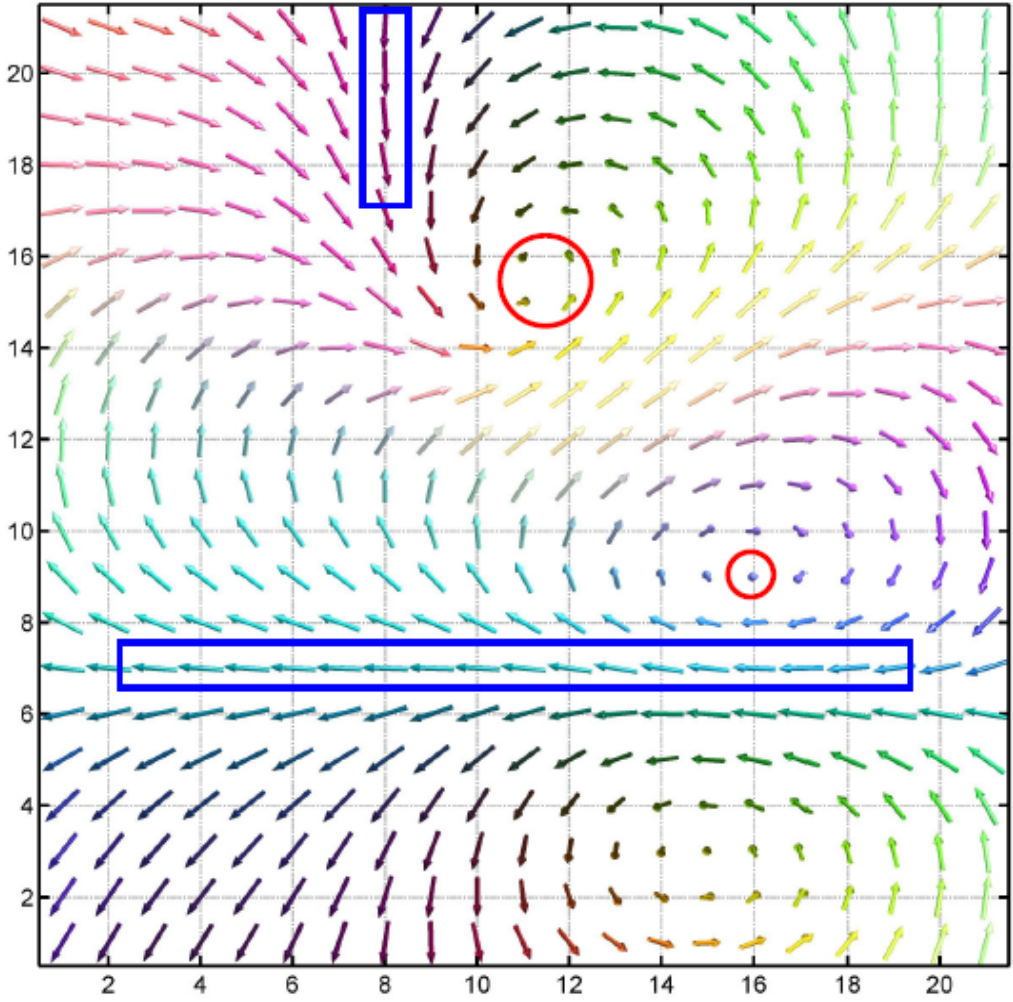

Figure S1. Cut through optimized simple cubic cluster illustrating the triple-twist spin order. Singularities (lines of parallel spins) in the plane of the cut are marked with blue rectangles. Red circles mark the ones running in the perpendicular direction (out of the page). Note the twist direction away from the singular lines (splay) in addition to the double-twist, which differentiates triple-twist from double-twist structures where the splay is absent.

*Experimental details*
We performed our experiments on the cold triple-axis spectrometer 4F and the SANS diffractometer PAPYRUS at the Orphée reactor of the Laboratoire Léon Brillouin, CEA, Saclay (France). On 4F we measured a 1-cm$^3$ single crystal of MnSi having resolution-limited lattice mosaic spread. $k_i$ was fixed at 1.15 Å$^{-1}$ with collimations open-20'(between double monochromator crystals)-20'-20' to obtain an energy resolution $\Delta E$ = 30 µeV [full width at half maximum]. To gain intensity at higher $T$, $k_i$ was 1.5 Å$^{-1}$ with collimations open-open-40'-20', yielding $\Delta E$ =145 µeV.